# Did Alice Do Wrong? Cross-Cultural Differences in Student Perceptions of Generative AI Use in University Computing Education

BRIAN HARRINGTON, University of Toronto Scarborough, Toronto, Ontario, Canada
IRINA ZLOTNIKOVA, School of Pure and Applied Sciences, Botswana International University of Science and Technology, Palapye, Botswana
GAYATHRI NADARAJAN, Sungkyunkwan University, Jongno-gu, Seoul, Korea (the Republic of)
SAMUEL EKUNDAYO, Open Polytechnic of New Zealand, Lower Hutt, New Zealand

The rise of generative AI (GenAI) in higher education has prompted urgent debates surrounding academic integrity and ethical use. This study examines cross-cultural differences in student perceptions of GenAI use, comparing responses from students at Canadian and South Korean universities. Using a scenario-based survey administered in Fall 2024, we analyzed how students judged the ethicality and rule compliance of AI-assisted coding practices. Results reveal that Canadian students were consistently more likely to perceive the use of GenAI as both unethical and against institutional policies compared to Korean students, despite functionally identical institutional policies. Statistical analysis, including Mann-Whitney U tests and correlation coefficients, demonstrated significant differences across nearly all scenarios. Analysis of the factors used in generating scenarios indicated that the amount of AI-generated code incorporated into assignments most strongly influenced ethical judgments. Findings were interpreted through Hofstede's cultural dimensions framework, suggesting that cultural factors such as power distance, individualism, and uncertainty avoidance significantly shape students' ethical reasoning regarding GenAI. Our results contribute to the growing body of evidence emphasizing that equitable AI integration in education must be culturally responsive, taking into account diverse conceptions of academic integrity. We advocate for the development of nuanced AI-use guidelines that are sensitive to local cultural contexts while upholding fundamental principles of academic honesty. This study highlights the need for ongoing cross-cultural research to inform ethical AI policies and support responsible GenAI use in global higher education settings.

This work began as part of the ACM Global Computing Education (ACM CompEd 2025) Conference's Partnership Projects Initiative. We would like to thank Michelle Craig (University of Toronto, Canada) and Samuel Mann (Otago Polytechnic, New Zealand) for their work on this fantastic program, without which this research project would never have gotten off the ground.
Authors' Contact Information: Brian Harrington (corresponding author), University of Toronto Scarborough, Toronto, Ontario, Canada; e-mail: brian.harrington@utoronto.ca; Irina Zlotnikova, School of Pure and Applied Sciences, Botswana International University of Science and Technology, Palapye, Botswana; e-mail: zlotnikovai@biust.ac.bw; Gayathri Nadarajan, Sungkyunkwan University, Jongno-gu, Seoul, Korea (the Republic of); e-mail: gaya@g.skku.edu; Samuel Ekundayo, Open Polytechnic of New Zealand, Lower Hutt, New Zealand; e-mail: Samuel.Ekundayo@openpolytechnic.ac.nz.

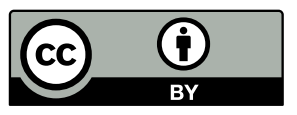

## 1 Introduction

The emergence of **generative AI (GenAI)** tools has led to both excitement and concern within higher education worldwide. Although educators and institutions acknowledge its potential to improve learning efficiency, provide personalized academic support, and assist with writing, ongoing debates highlight issues related to academic integrity, over-reliance on AI, and its implications for students' critical thinking skills [9, 59]. Some scholars view GenAI as a transformative and disruptive force, reshaping traditional knowledge production and assessment frameworks, thereby challenging conventional educational practices [59]. Furthermore, ethical and social concerns such as bias, fairness, and potential job displacement emphasize the need for robust regulatory frameworks and institutional oversight [62]. The risk that students may leverage AI to generate work that is not their own raises pressing questions about the authenticity of learning outcomes and the future role of AI in academic evaluation [9].

Cultural factors significantly shape students' engagement with GenAI in higher education, influencing both its adoption and ethical perceptions through community norms, technological infrastructure, and institutional policies [13]. According to Essien et al. [13], attitudes toward technology, academic integrity, and the alignment of AI with educational values vary across cultural contexts, directly affecting how students perceive GenAI as a learning tool and what they define as ethical use. In the Global South, cultural and structural challenges including limited access to AI, inadequate institutional training, socioeconomic disparities, and unclear governance of AI contribute to greater hesitancy, increased concerns about academic dishonesty, and inconsistent AI usage compared to the more AI-integrated educational systems of the Global North.

Students' and educators' views on what constitutes ethical use of AI in coursework can vary widely across cultural contexts. Western universities (e.g., in Canada and the US) have long emphasized individual originality and strict anti-plagiarism norms [3, 6, 58]. In contrast, East Asian educational cultures (e.g., in China and Korea) may approach academic integrity from different traditions, often influenced by Confucian principles that prioritize collective knowledge and respect for authoritative texts [6, 45]. Plagiarism perceptions vary across cultures, and the academic and professional worlds often have different attitudes toward it, with collectivist cultures sometimes viewing AI-assisted work as a form of collaborative learning rather than academic misconduct [29, 68].

In light of these challenges, our study investigated how students from Canada and South Korea perceived the ethicality and rule compliance of GenAI use in coursework. The **Research Objectives (ROs)** were as follows:

- *RO1*: To examine and compare students' perceptions of the ethical and institutional dimensions of GenAI use in coursework across Canada and South Korea.
- *RO2*: To analyze how specific scenario factors, such as the extent of code copied, level of understanding demonstrated, and the stage at which GenAI tools were used, influence students' ethical judgments in each cultural context.
- *RO3*: To interpret cross-cultural differences in students' perceptions and ethical evaluations using Hofstede's cultural dimensions framework.

The remainder of the article is structured as follows. Section 2 reviews related work and theoretical foundations, Section 3 describes the survey design and experimental methodology, Section 4 presents the results of the experiments and provides statistical analysis of the data, Section 5 discusses the findings, and Section 6 concludes with study contributions, implications for policy and practice, limitations, and future research directions.

## 2 Background

In this section, we explore four interconnected themes that directly support the study's ROs:

(1) Cross-cultural differences in perceptions of academic integrity, with a particular focus on East Asian and Western perspectives. This theme supports both RO1 and RO2 by framing the comparative aspect of students' ethical evaluations—see Section 2.1.
(2) Ethical concerns associated with GenAI in higher education and, specifically, computing education, which contextualize RO1 by highlighting the normative and institutional debates surrounding AI use in higher education—see Section 2.2.
(3) Insights from recent studies on how students from different cultural backgrounds perceive the ethical implications of GenAI, which inform RO2 by identifying key factors that impact ethical judgments—see Section 2.3.
(4) Hofstede's theory of cultural dimensions, which underpins RO3 by providing a theoretical basis for interpreting cross-cultural differences in student attitudes toward GenAI use—see Section 2.4.

Drawing on peer-reviewed research published primarily between 2023 and 2025, a time frame reflecting the emergence of GenAI, this analysis synthesizes key findings within thematic sections, critically evaluates their implications, and identifies gaps that warrant further exploration in future research.

### 2.1 Cultural Variations in Academic Integrity Perception

Academic dishonesty is viewed and tolerated differently across cultures, shaped by underlying social values and educational norms. This subsection synthesizes cross-cultural differences in perceptions of academic integrity, with attention to how these cultural frames may influence responses to GenAI use.

In collectivist cultures such as Lebanon, the United Arab Emirates, China, and the Philippines, certain forms of academic misconduct—particularly peer assistance—are often interpreted as morally acceptable due to cultural expectations of group loyalty, solidarity, or shared knowledge. For instance, students in these contexts may consider helping peers during exams or assignments as an ethical obligation rather than misconduct [33, 53, 61]. McCabe et al. [34] found that Lebanese students reported higher levels of academic dishonesty than their American counterparts, attributing this to cultural norms that prioritize collaborative problem-solving over individual performance. Likewise, students from the UAE exhibited more academic misconduct than American students, reinforcing the idea that collectivist cultural upbringing influences perceptions of integrity [61].

Privitera [45] highlights that Chinese education systems historically display greater tolerance toward plagiarism, often due to limited institutional training in citation practices. In the Philippines, Sonajo [53] shows how students use neutralization strategies, such as peer loyalty and external pressures, to rationalize dishonesty, further underscoring cultural norms of interdependence.

In contrast, individualistic cultures—such as those in Canada, the United States, and other Western contexts—tend to emphasize personal achievement, originality, and strict adherence to anti-plagiarism standards [6]. Students in these settings are socialized to regard academic work as individual intellectual property, where unauthorized collaboration or failure to cite constitute unethical behavior. This value system reinforces the idea that integrity lies in independent, original contributions.

East Asian education systems, including those of South Korea, Japan, and China, occupy a unique position. While generally collectivist, they emphasize respect for authority, rote memorization, and reliance on established sources, which influences how students perceive ownership of knowledge [6]. In these systems, knowledge may be viewed as communal rather than individually generated, contributing to differing interpretations of plagiarism and GenAI use.

As mentioned earlier, helping peers in Middle Eastern contexts is sometimes seen as a moral duty rather than cheating, reflecting broader communal expectations [14]. Similarly, in South East Asian contexts—such as within Filipino cultural values–helping peers is regarded as a virtue rather than a violation of integrity standards [53].

However, Simon [52] cautions against overgeneralizing these cultural patterns. While cultural norms play a role, she argues that issues of academic misconduct often arise from inadequate institutional support, insufficient training in academic writing, and language barriers, rather than inherent cultural values.

Taken together, these studies reveal that cultural context significantly shapes how students interpret and respond to academic integrity norms. This has important implications for understanding GenAI use across different educational systems, particularly when evaluating ethical judgments through a cross-cultural lens. However, the focus on collectivist versus individualist reflects only one dimension of cultural variation. Additional cultural dimensions relevant to academic integrity—such as power distance and uncertainty avoidance—are examined in Section 2.4.

## 2.2 Academic Integrity Concerns in GenAI Use

*2.2.1 General Academic Integrity Concerns in GenAI Use in Higher Education.* GenAI has rapidly emerged as a "double-edged sword" in academia, offering both transformative benefits and serious ethical challenges [10, 49]. Educators acknowledge the positive impact of **Large Language Models (LLM)** like ChatGPT, including increased student engagement, personalized feedback, and improved accessibility to learning resources [10]. However, these tools also undermine traditional academic integrity and assessment frameworks, raising concerns about their potential for plagiarism and misconduct. One of the most pressing issues is the ease with which students can generate answers using AI and submit them as their own work, blurring the line between assisted learning and academic dishonesty [10]. As pointed out in [47], ChatGPT behaves as an outstanding student across some of the modules, but also it performs consistently as an "adequate" student able to pass assessments without drawing undue attention to themselves.

Compounding this problem is the fact that AI-generated content can evade traditional plagiarism detection software, sparking fears of an "arms race" between AI-facilitated cheating and AI-powered detection technologies [10, 60]. Early studies following ChatGPT's release have already reported instances of AI-assisted dishonesty in coursework, further amplifying institutional concerns [38]. Another key ethical dilemma involves authorship and originality. The submission of AI-generated text challenges the fundamental concept of student authorship, with some scholars labeling this phenomenon "AI-giarism"—a term referring to plagiarism facilitated by AI [7]. Universities are still grappling with how to define and regulate GenAI use, debating whether its undisclosed application constitutes unauthorized collaboration or a violation of honor codes.

Despite these challenges, institutions are not powerless in managing GenAI's impact. Bobula [4] suggests that universities can harness the benefits of LLMs while mitigating risks through strategies such as updating assessment designs, providing AI literacy training, and implementing clear usage guidelines. Rather than resorting to outright bans, scholars advocate for a structured and ethical approach to AI integration. Cotton et al. [10] recommend that universities establish

AI-use policies, student and faculty training programs, and improved cheating detection methods to embrace AI's opportunities while maintaining academic integrity. ChatGPT is described as both an enabler of learning and a potential threat to academic honesty, prompting institutions worldwide to redefine plagiarism policies, honor codes, and assessment methods in response to GenAI's capabilities [10, 60].

The general consensus is that GenAI is here to stay. As Cotton et al. [10] conclude, universities must adapt proactively by establishing clear guidelines and innovating pedagogical strategies to preserve academic integrity in an AI-enhanced educational landscape.

*2.2.2 Academic Integrity Concerns in GenAI Use in Computing Education.* There are also academic integrity concerns that are specific to computing education, particularly due to the discipline's emphasis on programming assignments, code generation, and algorithmic problem-solving—tasks that GenAI tools can perform with high accuracy.

Lau and Guo [28] captured educators' evolving attitudes toward AI code-generation tools like ChatGPT and Copilot in programming modules. The study demonstrated a shift from an initial resistance or total ban to gradual acceptance, adaptation, and proactive integration of these tools into programming education.

Pang and Vahid [41] note that cheating remains a persistent concern in introductory computer science courses, especially among students with limited programming experience. The study by Prather et al. [44] highlights that students' use of GenAI tools in computing education raises serious academic integrity concerns, particularly when AI-generated code is submitted without proper understanding or attribution. Such practices undermine learning outcomes in programming courses, where developing algorithmic thinking and problem-solving skills is essential. Additionally, the authors emphasize the difficulty in assessing student competence fairly and maintaining clear boundaries of authorship in GenAI-assisted programming tasks.

Padiyath et al. [40] found that early use of GenAI tools in a programming course was linked to lower self-efficacy and weaker performance, especially when students relied on AI rather than using it strategically. Their study highlights how peer influence and career goals shape tool adoption, raising academic integrity concerns beyond plagiarism—such as reduced learning autonomy.

Similarly, the study conducted across Kenya, Nigeria, and South Africa specifically investigated academic integrity concerns in computing education by examining how students in computer science and programming courses interact with GenAI tools [39]. While GenAI tools can provide individualized feedback and enhance access to programming support, they also risk promoting superficial engagement and academic shortcuts, particularly among novice learners unfamiliar with foundational concepts. Students expressed uncertainty about the legitimacy of using AI-generated code in assessments, raising critical questions around authorship, responsible use, and the need for institutional guidance to uphold academic integrity in AI-assisted programming tasks [39].

A mini-review synthesizing empirical studies on ChatGPT in programming education reported that while many students found GenAI helpful for debugging and understanding code, there are widespread concerns about misuse, dependency, and potential erosion of core skills [11]. These concerns directly relate to academic integrity, as students may bypass genuine learning by overly relying on AI-generated solutions.

Humble [20] provides a broader thematic analysis of GenAI in computing education and outlines ethical threats such as skill dilution, misuse, and lack of transparency in authorship. The study presents a risk management strategy that emphasizes the importance of institutional safeguards and pedagogical interventions to mitigate academic integrity violations.

Finally, Ye et al. [63] conducted a quasi-experimental study with students learning programming through GenAI-enhanced tools. While students reported improved self-efficacy, the study cautioned that unguided GenAI use may lead to surface-level learning and ethical ambiguity in task ownership. The authors call for structured integration and clear ethical guidance to ensure that academic integrity is preserved in GenAI-assisted computing education.

It should be noted that only two papers [39, 44] incorporate findings from several countries. However, the article by Prather et al. [44] does not specifically examine cross-cultural perceptions of GenAI use. The rest of the reviewed literature on GenAI and academic integrity in computing education is limited to single-country contexts, remaining largely context-specific and lacking an explicit cross-cultural perspective.

### 2.3 Cross-Cultural Student Perceptions of GenAI Ethics

In Canada, researchers have begun to explore cultural perspectives on GenAI ethics, although the studies remain limited. Kumar [27] found that Canadian students, along with peers in other Western, English-speaking countries, generally perceived GenAI use as increasingly permissible over time. This suggests a relatively high acceptance of AI tools in academic settings. The study emphasized variability in ethical views depending on academic discipline and institutional culture. Similarly, Kim et al. [25] and Stone [55] reported that Canadian and US students showed discipline-specific ethical concerns: STEM students tended to view GenAI as a legitimate tool for innovation, while humanities students emphasized authorship and originality. These studies highlighted Canada's individualistic academic culture, which values personal accountability and is shaped by institutional policies more than collective ethical norms.

In South Korea, students experience significant academic pressure, which influences their perceptions of GenAI. Some regard ChatGPT as a helpful support tool, while others fear over-reliance could erode their problem-solving skills [42]. The education system's emphasis on independent learning creates tension with AI assistance. Students were divided over the ethics of using GenAI: some saw it as acceptable with transparency, others rejected it entirely [64]. Korean students also reported difficulties with ChatGPT's performance in Korean, often preferring responses in English [37, 65]. This tendency led them to rely more heavily on authoritative sources, reflecting the cultural norms and practices that shape AI adoption in education.

In China, academic integrity is defined by a hierarchical education system and high-stakes testing. GenAI is often seen as an extension of collective knowledge rather than a violation of personal ethics [68]. Government regulation plays a significant role, with students expecting institutional validation of AI content [29]. AI tools are commonly used for collaborative learning and efficiency, not seen as inherently unethical [59]. However, concerns remain about linguistic authenticity and potential misuse in writing tasks [9].

In Hong Kong, Chan [7] found students clearly understood traditional plagiarism rules but were less certain about GenAI. The concept of "AI-giarism" emerged, reflecting difficulty in applying existing norms to new technologies. Students often deferred to university policies rather than personal judgment, aligning with a broader East Asian tradition of institutional authority and structured ethics.

Elsewhere in Asia, Chou [8] found that faculty across China, Japan, South Korea, and Taiwan held culturally distinct views on GenAI ethics. Japanese and Korean scholars emphasized authorship and integrity, while Chinese respondents focused on regulatory compliance. Taiwanese perspectives were pragmatic, valuing transparency and educational enhancement.

In Africa, a study across Kenya, Nigeria, and South Africa revealed diverse attitudes toward GenAI in programming education [39]. Students in Nigeria and Kenya viewed AI tools as beneficial for access and equity, whereas South African students expressed more skepticism, shaped by better infrastructure and greater exposure to digital tools.

In Tunisia, Kamoun et al. [24] showed how financial constraints, ambiguous policies, and institutional authority shaped students' ethical reasoning. Many students justified AI-assisted plagiarism under specific conditions, navigating ethics pragmatically.

In two European countries (Serbia and Austria), Adžić et al. [1] found that Serbian students were cautious toward GenAI, reflecting a collectivist and economically constrained context, while Austrian students, influenced by individualism and technological optimism, embraced GenAI for innovation and learning.

In Sweden, a large-scale study by Malmström et al. [32] revealed that students largely supported regulated GenAI integration. They distinguished between legitimate educational use and unethical behaviors like plagiarism, consistent with Sweden's emphasis on digital literacy and trust-based education systems.

Across Anglophone Western countries (Australia, New Zealand, UK, US), Kumar [27] found variation in GenAI ethics based on national and disciplinary cultures. For example, UK students expressed strong concern about AI undermining creativity, while US students debated AI's role in authorship and accountability.

To better understand variations in ethical perceptions of GenAI by students across various regions and interpret the emerging patterns in a structured way, the following subsection draws on Hofstede's cultural dimensions theory. Hofstede's theory offers a theoretical lens to analyze these cross-cultural differences.

### 2.4 Hofstede's Theory as a Basis for Interpreting Cross-Cultural Differences in GenAI Perceptions

To systematically explain cross-cultural variations in perceptions of academic integrity and, specifically, GenAI ethics, researchers frequently invoke Hofstede's cultural dimensions theory [33, 58, 66, 67]. This framework identifies key dimensions along which national cultures differ [16–18]. The dimensions of power distance, individualism versus collectivism, competitiveness ("masculinity vs. femininity" as initially formulated by Hofstede), and uncertainty avoidance are particularly useful in explaining variations in academic dishonesty across cultural contexts.

Several other major frameworks have also been developed to explain cross-cultural variation, including Schwartz's theory of basic values [50], the **World Values Survey (WVS)** [22], Trompenaars' Seven Dimensions of Culture [57], the GLOBE study [19], and Schwartz's national-level cultural value orientations [51]. While Schwartz's and the WVS frameworks provide valuable insights into broad motivational and societal value patterns, they primarily focus on individual priorities or macro-level sociopolitical trends, making them less directly applicable to the educational context of this study. Similarly, Trompenaars' model and the GLOBE study emphasize cultural influences on leadership and organizational behavior, offering limited relevance for understanding students' ethical judgments within academic institutions. In contrast, Hofstede's model provides a direct explanatory basis for how national culture shapes individual behavior in structured settings like universities, particularly regarding perceptions of authority, rule-following, and uncertainty around emerging technologies like GenAI. Consequently, Hofstede's cultural dimensions offer the most precise and contextually appropriate theoretical foundation for analyzing cross-cultural differences in students' ethical evaluations of GenAI use in university education.

In high power distance cultures, authority figures (e.g., professors) are greatly respected, and students may refrain from questioning rules, waiting instead for explicit instructor guidance on GenAI usage [33]. If GenAI use is not explicitly prohibited, students in such cultures may assume its use is acceptable. In contrast, low power distance cultures encourage open dialogue between students and faculty, making students more likely to seek clarification on AI policies and to report observed academic dishonesty.

Individualistic cultures (e.g., Canada and the US) prioritize personal achievement and originality, making students more likely to view AI-generated content as a violation of academic integrity [6]. In contrast, collectivist cultures such as those in East Asia might perceive GenAI as a shared resource that facilitates group success, aligning with traditional knowledge-sharing practices rather than being seen as unethical [6]. This aligns with Hofstede's perspective that collectivist societies emphasize group goals over individual originality, potentially making students less likely to think in terms of "I" and proprietary ideas [6]. Similarly, Marhoon and Wardman [33] found that New Zealand students, from an individualistic culture, viewed academic dishonesty as a hindrance to self-development, whereas Bahraini students, from a collectivist culture, considered cheating as an act of "helping" rather than dishonesty. Individualistic cultures tend to view cheating as an individual moral failing, whereas collectivist cultures perceive academic dishonesty as context-dependent, often influenced by situational or relational factors [30]. An individualist-oriented student may feel personal guilt about misusing GenAI, while a collectivist-oriented student may experience shame if their actions bring dishonor to their group or institution [26].

Academic competitiveness also influences how far students are willing to go to succeed [35, 43, 46]. In highly competitive academic environments, both in Eastern and Western contexts, the intense pressure to achieve top scores can drive some students to exploit GenAI tools as a strategic means to secure academic success [59]. This aligns with findings from computing education, where the most frequently cited motivation for plagiarism is the pressure to achieve good grades [2]. Conversely, in cultures that emphasize ethical considerations and fairness over competition, students may regulate their GenAI use more carefully.

The fourth dimension, uncertainty avoidance, which reflects how much a person or culture feels threatened by uncertainty and ambiguity, also influences attitudes toward GenAI adoption [58]. Cultures with high uncertainty avoidance (e.g., South Korea or Japan) tend to perceive GenAI tools as risky and potentially dishonest, preferring strict rules regarding their use [58]. According to Im et al. [21], students in high uncertainty avoidance cultures exhibit greater caution in adopting new digital technologies. Conversely, low uncertainty avoidance cultures (e.g., Denmark or Sweden) are more open to integrating GenAI into academic work as an innovative tool rather than an ethical threat [66]. Prather et al. [44] report data from a survey involving computing students and instructors across 20 countries. Their analysis reveals a significant relationship between national-level uncertainty avoidance and the perception of GenAI as academic misconduct. Similar to [58] and [66], findings of Prather et al. [44] suggest that individuals from high uncertainty-avoidance countries are more likely to view GenAI use as dishonest.

Based on Hofstede's cultural dimensions, Zhao et al. [67] conducted a meta-analysis synthesizing findings from 80 studies to examine how academic cheating behaviors are influenced by achievement orientations and cultural values, including comparisons of South Korea and Canada—which are the geographical scope of our empirical study. Using Hofstede's theory as theoretical framework, Zhao et al. [67] categorized Canada as a low power distance and highly individualistic country, whereas South Korea was classified as high power distance and collectivist. In high power distance cultures like South Korea, students may feel a stronger obligation to conform to social expectations and authority figures, leading to the rationalization of academic dishonesty as a necessary means to meet educational demands. In contrast, low power distance cultures like Canada emphasize individual accountability, fostering stricter adherence to academic integrity norms. Furthermore, the study found that individualistic cultures prioritize personal achievement and ethical responsibility, reducing tendencies toward academic dishonesty, whereas collectivist cultures emphasize group success, making certain forms of academic collaboration, viewed as misconduct in Western contexts, more acceptable.

## 3 Methodology

In order to evaluate student perceptions on the use of GenAI tools in computing education, we developed a survey instrument based on soliciting Likert scale responses to questions pertaining to a series of scenarios. The Likert scale is widely regarded as an appropriate instrument for measuring subjective phenomena such as attitudes, opinions, and perceptions, particularly within the fields of education, psychology, and social sciences [23].

### 3.1 Methodological Foundations of Scenario-Based Ethical Inquiry

The use of scenario-based instruments has become a well-established method in education research for eliciting nuanced ethical judgments. This approach draws on techniques from the social sciences that utilize short, hypothetical scenes—commonly referred to as vignettes or scenarios—to examine participant responses to ethically complex or context-sensitive situations [36]. These scenes are designed to mirror real-life dilemmas in a controlled manner, allowing researchers to explore how respondents reason through issues involving ambiguity, conflicting norms, or evolving technologies.

Several studies illustrate the effectiveness of this method. Russell et al. [48] used fictionalized newspaper articles to analyze how teacher characteristics influence public perceptions of misconduct, while Bourke et al. [5] employed scenarios to investigate ethical drift among educational psychologists. Şensoy and İkiz [69] demonstrated how scenario-based methods reveal decision-making strategies among school counselors in ethically ambiguous situations. In the domain of academic integrity, Mah et al. [31] used tailored scenario exercises to study student perceptions of ChatGPT use, revealing how context shapes judgments about cheating and learning. Similarly, Stephens and Bertram Gallant [54] applied moral dilemma vignettes to enhance students' moral sensitivity in response to academic misconduct. Finally, Möller et al. [36] advocated for bespoke scenarios to study student responses to third-party writing assistance, emphasizing how such tools enable deeper insight than traditional surveys.

These studies confirm that scenarios offer a flexible yet rigorous methodology for probing ethical perceptions in educational settings. They enable both cross-sectional comparisons and intra-population calibration, particularly when dealing with emerging issues like the ethical use of GenAI. Our study built on this body of work by designing and deploying custom scenarios to examine students' attitudes toward GenAI use in programming coursework.

### 3.2 Scenarios Used in the Survey

A total of eight scenarios were constructed with varying degrees of:

- Amount: Amount of generated code in final submission.
- Understanding: Amount of displayed understanding of generated results by student submitting the work.
- Stage: Stage at which GenAI tool was used.

Each scenario was coded on three dimensions—Amount, Understanding, and Stage—using a three-level ordinal scale (Low/Medium/High or Early/Middle/Late) defined in the codebook (Table 1). These definitions include decision rules that allow classification either by approximate percentage of the final code or, when percentage is impractical, by structural units (lines/functions). The scenario–factor assignments reported in Table 2 were derived directly from these operational definitions. The full text of the scenarios can be found in Figure 1.

Not all combinations were used; instead, we selected a representative subset that enabled meaningful comparisons while keeping the survey length reasonable. Scenarios were validated by asking two researchers to independently rank each scenario on the factors provided and ensuring that

Table 1. Codebook: Operational Definitions for Factors and Levels Used to Construct/Classify Scenarios

| **Factor** | **Low** | **Medium** | **High** |
|---|---|---|---|
| *Amount of AI-generated code in final submission* | None or only a few lines; trivial edits only. No complete function or block from AI is submitted. | A localized snippet integrated (e.g., one helper function or a short loop/method) with minor adaptation. One contained region replaced by AI output. | A substantial portion of the solution originates from AI (e.g., multiple regions or a core function) with little structural change. Multiple regions or a key function primarily from AI. |
| *Level of understanding demonstrated* | Copy/paste with superficial changes; no clear reasoning/debug evidence. Correctness via insertion, not explanation. | Some comprehension via limited edits or partial re-implementation; can explain "what" but limited understanding of "why/how." Modifies AI code to fit context or fixes obvious mismatches. | Independent reasoning: re-implements from principles, debugs/fixes AI output, explains logic/constraints. AI informs, but the final code is student-authored or substantially reworked. |
| *Stage of use* | *Early*: Before substantial coding (idea generation, scaffolding). | *Middle*: During development (filling a missing piece, writing a function). | *Late*: After a draft has been produced (debugging, verification, refactoring, documentation). |

Scenario texts appear in Figure 1.

Table 2. Scenarios Broken Down by Factors Used for Generation

| Scenario | Amount of generated code in final submission | Understanding demonstrated | Stage of use |
|---|---|---|---|
| Scenario 1 | Low | Low | Early |
| Scenario 2 | Low | Medium | Early |
| Scenario 3 | Low | High | Middle |
| Scenario 4 | High | High | Middle |
| Scenario 5 | Medium | High | Late |
| Scenario 6 | Low | High | Late |
| Scenario 7 | High | Low | Early |
| Scenario 8 | Low | Medium | Middle |

evaluations were consistent. As no disagreements were found, formal inter-rater statistics and resolution procedures did not apply.

To remove extraneous variables, assignment description, and description of the user of the tools were kept consistent. No details were provided as to the actual rules in either the course or institution in question.

Attached to each scenario was a pair of prompts:

Alice acted unethically in this scenario

**Scenario 1:** Alice is a student in an introductory programming course. For one of her assignments, she is unsure where to get started. She copies and pastes the text from the assignment handout into a generative AI tool and asks it to produce code. She reads the code to get a sense of how it works. She then closes the window without copying any code, opens her own IDE (code editor), and begins to write her own solution.

**Scenario 2:** Alice is a student in an introductory programming course. For one of her assignments, she is unsure where to get started. She copies and pastes the text from the assignment handout into a generative AI tool and asks it to produce code. She then copies the code into her IDE (code editor), and then, one function at a time, deletes the code from the function, and re-writes it on her own.

**Scenario 3:** Alice is a student in an introductory programming course. For one of her assignments, she is struggling with a particular function. She writes out the explanation for what she wants the function to do in a generative AI tool and asks it to produce code. She reads the code to get a sense of how it works. She then closes the window without copying any code, goes back to her own IDE and writes the function herself.

**Scenario 4:** Alice is a student in an introductory programming course. For one of her assignments, she is struggling with a particular function. She writes out the explanation for what she wants the function to do in a generative AI tool and asks it to produce code. She then copies 5 lines of code into her own function and makes minor edits to make it work with the code she has already written.

**Scenario 5:** Alice is a student in an introductory programming course. For one of her assignments, she is struggling with a particular function. She copies the code for her function into a generative AI tool, and writes a prompt explaining the bug that she cannot fix, asking for a working version of the code. She copies the produced code into her own function.

**Scenario 6:** Alice is a student in an introductory programming course. For one of her assignments, she is struggling with a particular function. She copies the code for her function into a generative AI tool, and writes a prompt explaining the bug that she can't fix, asking for a working version of the code. After reading the code produced by a generative AI tool, she understands the bug and fixes her own code

**Scenario 7:** Alice is a student in an introductory programming course. For one of her assignments, she is struggling with a particular function. She goes online to look for a solution to the problem, and she finds several recommendations. She copies each of them into a generative AI tool to know which of them works for her assignments. Upon confirming which code is correct, she copies the code into her own function.

**Scenario 8:** Alice is a student in an introductory programming course. For one of her assignments, she is struggling with a particular function. She goes online to look for a solution to the problem, and she finds several recommendations. She copies each of them into a generative AI tool to know which of them works for her assignments. Upon confirming which code is correct, she deletes it and tries to replicate what she has learned in her own IDE (code editor).

Fig. 1. The text of the eight scenarios presented to subjects.

and

> What Alice did would be against the rules in my institution

Participants were asked to respond on a five-point Likert scale ("Strongly Disagree," "Disagree," "Neither Disagree nor Agree," "Agree," and "Strongly Agree") to each prompt.

In addition to the scenario-based prompts, the survey asked for information about the year of study of the respondent, their familiarity with and usage of GenAI tools, and their overall perceptions of the impacts that GenAI is having in academia, with the following prompts:

> I often use GenAI tools in my university education.

> I believe that GenAI tools are having a negative impact on the academic integrity at my institution.

> I believe that GenAI tools are having a negative impact on the quality of learning at my institution.

> I believe that GenAI tools will improve fairness among students.

These prompts were intended to evaluate whether there was any higher-level bias towards GenAI tools between the populations being studied. These bias-assessment items were included in the same survey instrument as the scenarios. The bias questions were administered after the scenario-based responses in order to avoid influence that could be caused by meta-cognition on personal bias.

### 3.3 Survey Implementation

The survey was administered at two separate institutions in Fall 2024. University of Toronto Scarborough (hereafter referred to as *CAN*) is a large, North American research intensive institution. At CAN, the surveyed course was an elective in introductory programming for students not majoring in computer science. This did not affect generalizability of findings because we targeted non-major introductory programming, where curriculum and assessment were comparable whether a course was required or elective. Our mixed-discipline sample was typical of such cohorts. Sungkyunkwan University (hereafter referred to as *KOR*) is a large, East Asian research intensive institution. The courses surveyed at KOR were three data science-related courses; relational databases, natural language processing and ethical and social impact of data, although not all students were data science majors.

Both CAN and KOR had similar institutional context in relation to the use of GenAI at the time in question, in that there were no clear institution-wide mandates, and decisions about acceptance or use of GenAI tools was left to the discretion of the instructor. In both contexts surveyed, the policy was identical. In both institutions, the survey was administered at the end of the course, but before the final exam. Participation in the survey was voluntary. Students could earn a small participation bonus for completing the survey. However, consent to use their responses for research was optional and had no effect on the bonus.

## 4 Results

A total of 262 students at CAN (course size 420, response rate 62%) and 48 students at KOR (course size 65, response rate 74%) completed the survey and agreed to the use of their data for research. Unequal sample sizes reflected enrollment differences; we surveyed all available students during the study period. Expanding the KOR sample was not feasible, and adding data later would compromise comparability given the elapsed time and evolving perceptions of GenAI.

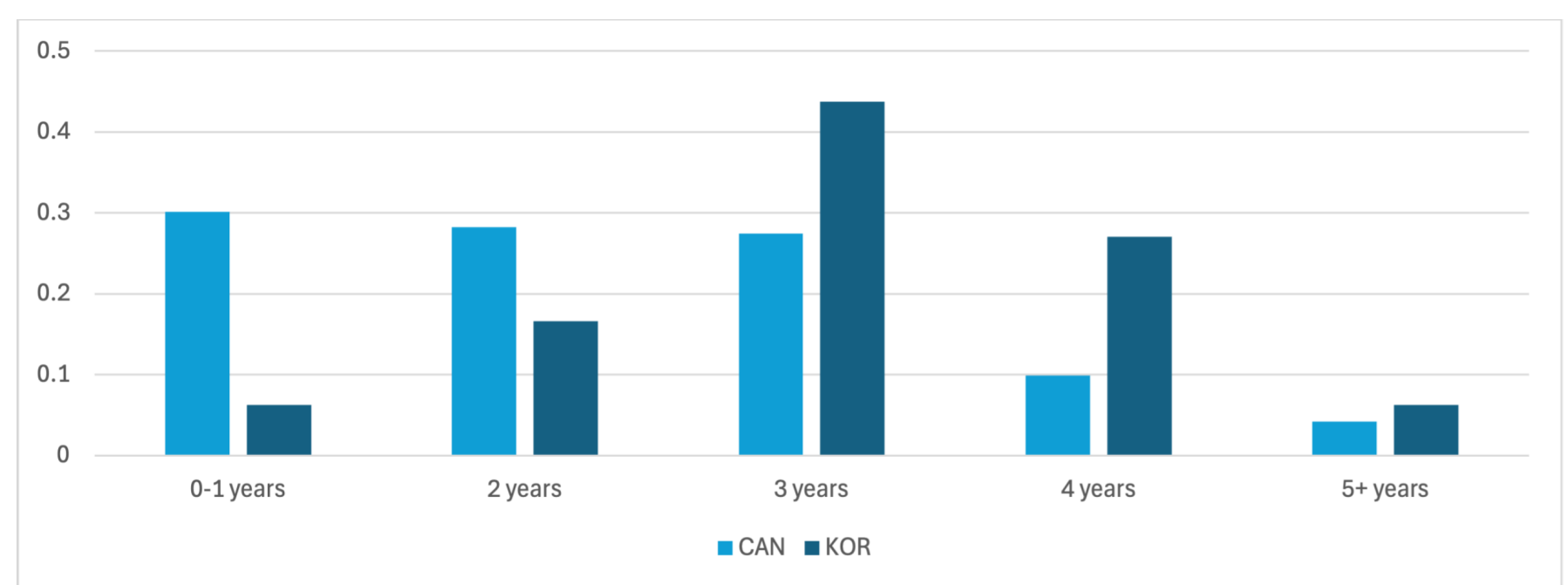

Fig. 2. Academic progress distribution in surveyed populations.

Table 3. Two-Way ANOVA Results for Each Judgment Outcome

| Source of variation | Sum of squares | df | Mean square | F | p |
|---|---|---|---|---|---|
| Country (“Acted unethically”) | 27.5412 | 1 | 27.5412 | 71.0336 | **<0.0001** |
| Year of study (“Acted unethically”) | 1.3314 | 4 | 0.3329 | 0.8585 | 0.4891 |
| Interaction (“Acted unethically”) | 0.0959 | 4 | 0.0240 | 0.0618 | 0.9929 |
| Country (“Against the rules”) | 41.5414 | 1 | 41.5414 | 85.8967 | **<0.0001** |
| Year of study (“Against the rules”) | 1.82 | 4 | 0.455 | 0.9408 | 0.4406 |
| Interaction (“Against the rules”) | 0.323 | 4 | 0.08076 | 0.167 | 0.9999 |

Factors: Country and year of study with the Country × Year interaction; dependent variables: “Acted unethically” and “Against the rules at my institution.” Bold indicates significance after Bonferroni-corrected $\alpha = 0.025$.

In this section we first assess whether other factors (academic progress, overall opinion on educational impact of GenAI) were shown to impact the results, we then analyze the prompt responses between the two student populations and attempt to discover which factors of the questions had the largest impact on difference in responses between the groups.

## 4.1 Academic Progress

Students in KOR were slightly further along in their studies than those in CAN. As can be seen in Figure 2, the majority of students in CAN took the course in their first two years of study, whereas students in KOR mostly took the course in the third and fourth years of study. This is a natural consequence of the academic calendars of the institutions in question.

To assess whether this might affect our results, we performed a two-way ANOVA on the Likert scale responses for each prompt type, with factors Country and Year of Study, across the full dataset. As shown in Table 3, the academic progress was not a significant factor for either judgment outcome (“Acted unethically” $p = 0.4891$ or “Against the rules” $p = 0.4406$), with very little interaction effect ($p > 0.99$ in both categories).

## 4.2 Overall Opinion on Educational Impact of GenAI

The overall responses to the prompts regarding the impacts of GenAI tools on academia were found to be very similar between groups. As can be seen in Figure 3, despite students in CAN being more likely to respond neutrally than those in KOR, the overall distribution was very similar. As shown in Table 4, Mann–Whitney U tests showed no statistically significant difference in responses between the groups for either prompt.

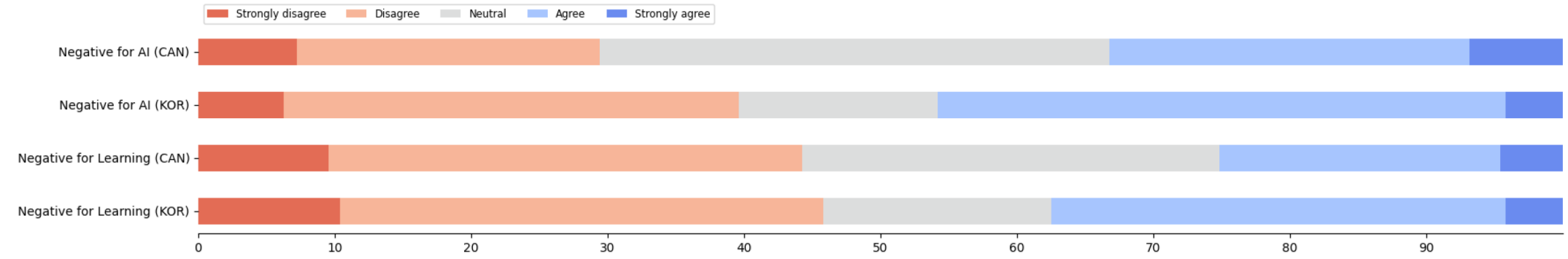

Fig. 3. Responses to prompts regarding impact of GenAI on academia: "I believe that GenAI tools are having a negative impact on the academic integrity of students in my degree" (top two) and "I believe that GenAI tools are having a negative impact on the quality of learning of students in my degree" (bottom two).

Table 4. Overall-Opinion Items: Mann–Whitney U Test Results ($U$, $z$, Effect Size, p)

| Scenario and prompt | $U$ | $z$-score | Effect size | p |
|---|---|---|---|---|
| Negatively impacting academic integrity | 6,201.5 | −0.15 | 0.0085 | 0.88 |
| Negatively impacting learning | 5,987 | −0.53 | 0.0301 | 0.59612 |

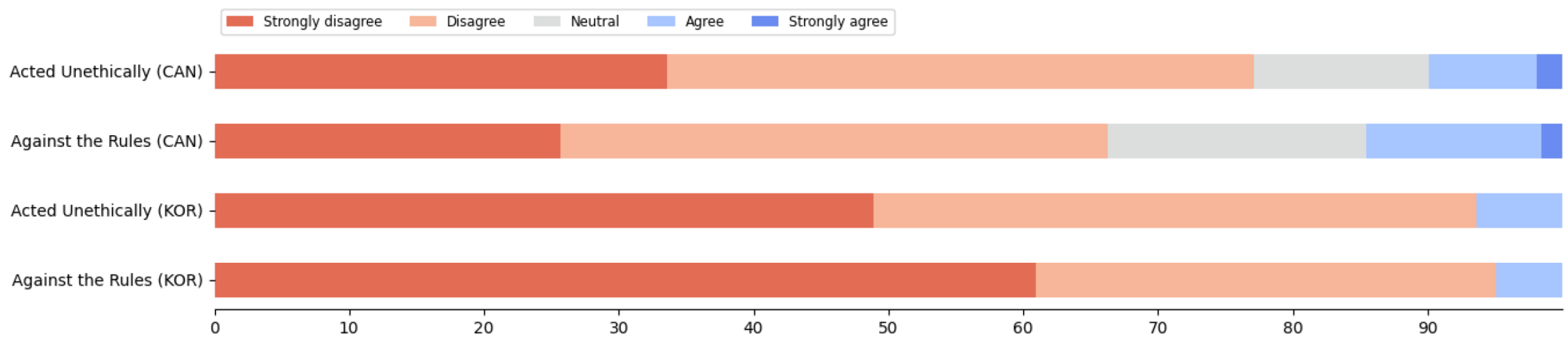

Fig. 4. Responses to Scenario 1. Amount: Low; Understanding: Low; Stage: Early.

### 4.3 Perceptions of Ethicality and Institutional Rules

There was a clear pattern in the student responses across all scenarios. Students at CAN were much more likely to say that Alice acted unethically (mean = 2.6 on five-point Likert scale) than students at KOR (mean = 1.78), and that what Alice did would be against the rules at their institution (CAN mean = 2.81, KOR mean = 1.80). This was despite the actual rules at the two institutions being functionally identical.

Given the ordinal Likert responses, we used Mann–Whitney U tests to compare CAN and KOR on each item. An overall rank-based comparison across the eight scenario items showed statistically significant differences in both ethicality ($z = 7.15$, $p < 0.00001$) and rule-compliance ($z = 7.43$, $p < 0.00001$). On each individual question, the same pattern occurred, with all prompts showing statistically significant differences except for Scenario 1: Alice acted unethically in this scenario. Summary results can be found in Table 6, and individual prompt responses are shown in Figures 4–11 (the full descriptions of the scenarios are available in Figure 1). For readers preferring a descriptive view, the scenario-level distributions in these figures can be read visually. We summarized these patterns descriptively and made no additional inferential claims beyond the reported Mann–Whitney tests.

Across the eight scenario items, the response distributions differed systematically between the two countries. Because the Mann-Whitney U test compares ranked distributions and does not require identical shapes, we interpreted significant results as distributional (ordinal) differences rather than strictly median differences [12]. We did not attribute these contrasts solely to underlying ethical judgements. Potential response-style differences (e.g., cross-cultural variation in neutral-category use) may also contribute, so we refrained from group-specific attributions regarding neutral-option use in the scenario items. Readers who accept that between-country contrasts primarily reflect substantive ethical-judgement differences may interpret the Mann-Whitney U test results accordingly. Readers who do not (or who consider response-style explanations plausible) should treat these results with caution.

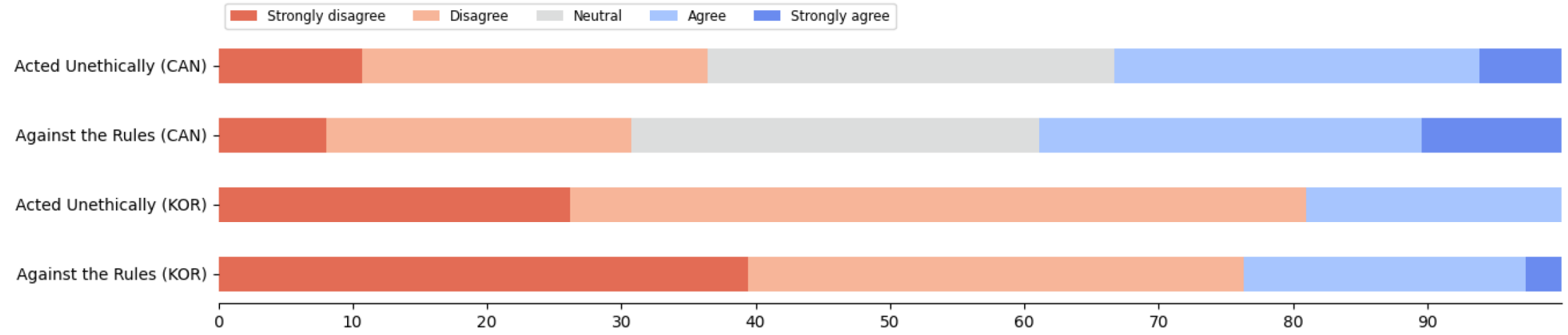

Fig. 5. Responses to Scenario 2. Amount: Low; Understanding: Medium; Stage: Early.

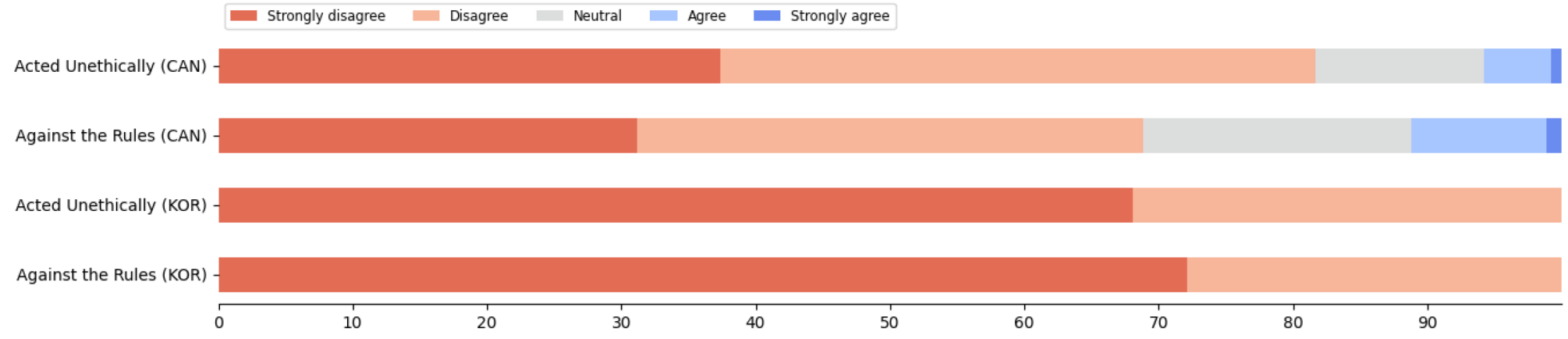

Fig. 6. Responses to Scenario 3. Amount: Low; Understanding: High; Stage: Middle.

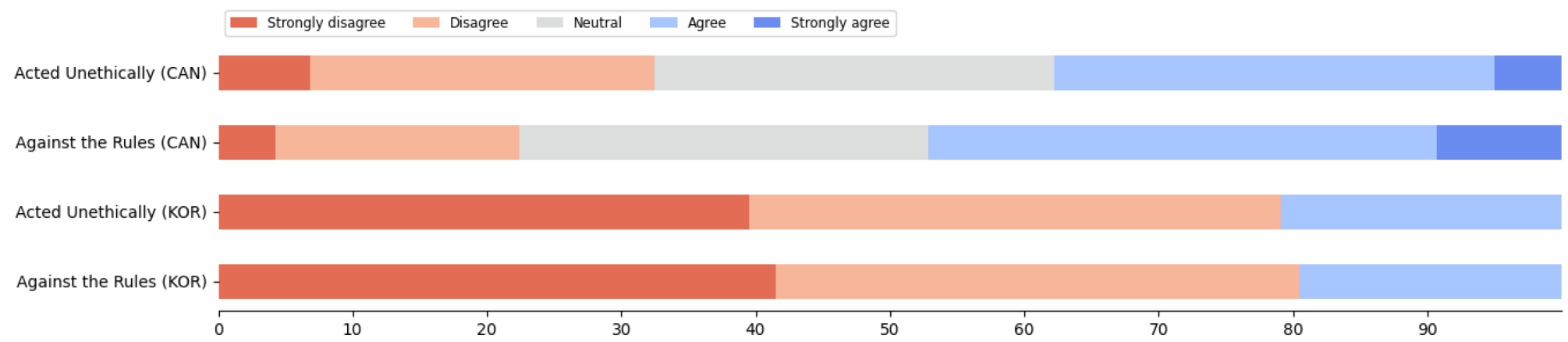

Fig. 7. Responses to Scenario 4. Amount: High; Understanding: High; Stage: Middle.

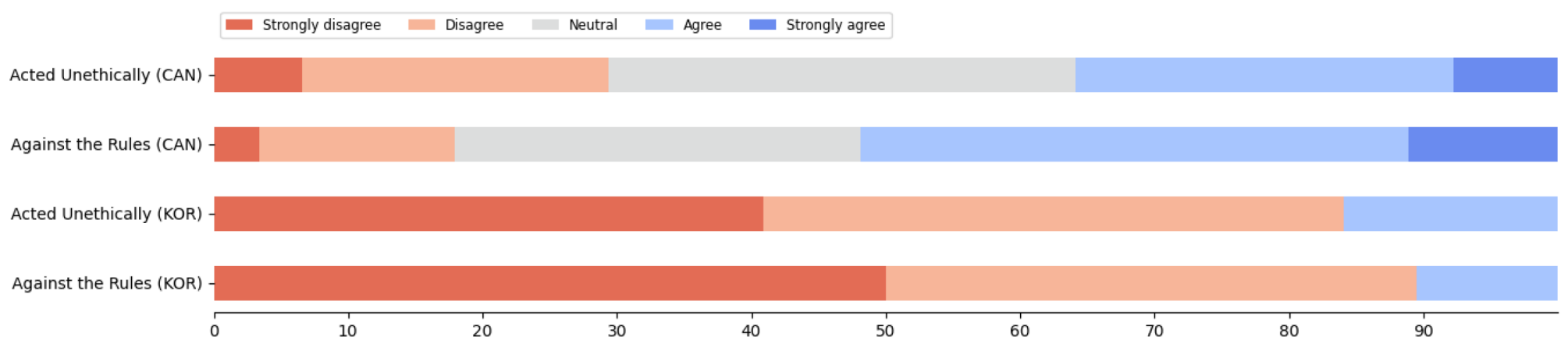

Fig. 8. Responses to Scenario 5. Amount: Medium; Understanding: High; Stage: Late.

Across the eight scenarios, the mean "Acted unethically" rating was consistently lower than the mean "Against the rules at my institution" rating as evidence by a paired $t$-test ($t(295) = -7.9879$, p =< 0.0001). A Pearson's correlation coefficient was used to measure the strength and direction of the linear association between students' perceptions of ethicality and rule compliance [15] showing a strong and statistically significant correlation between a student's perception of whether Alice acted unethically, and whether her actions would be against the rules at their institution ($r(294) = 0.8524$, $p < 0.0001$).

To assess the impact of the study factors on participant perceptions, we conducted a repeated-measures ANOVA that accounted for the repeated responses each participant provided across the

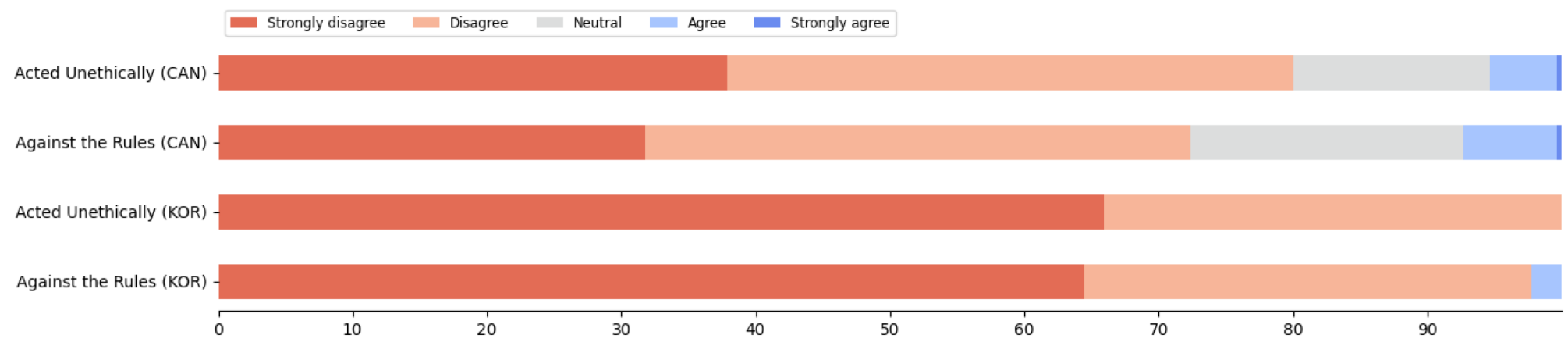

Fig. 9. Responses to Scenario 6. Amount: Low; Understanding: High; Stage: Late.

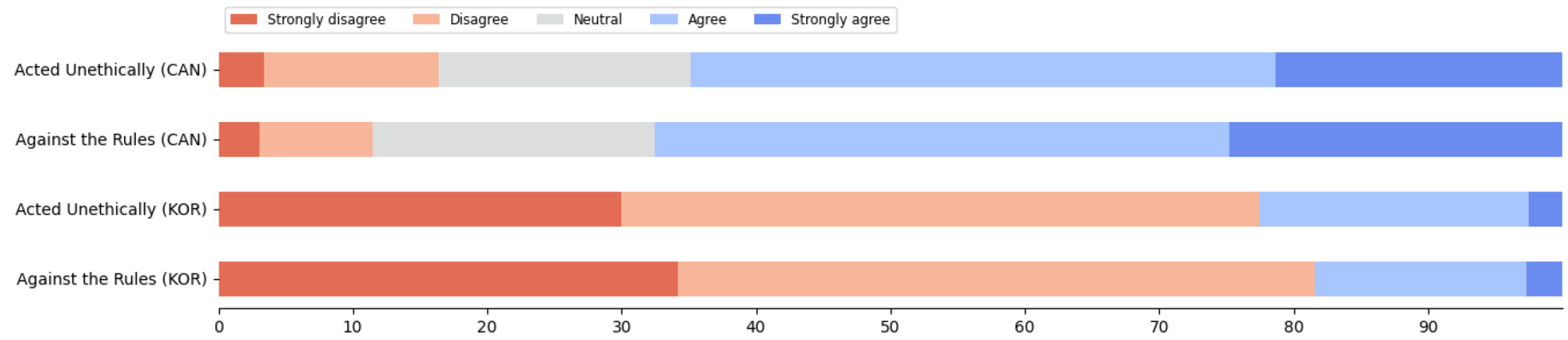

Fig. 10. Responses to Scenario 7. Amount: High; Understanding: Low; Stage: Early.

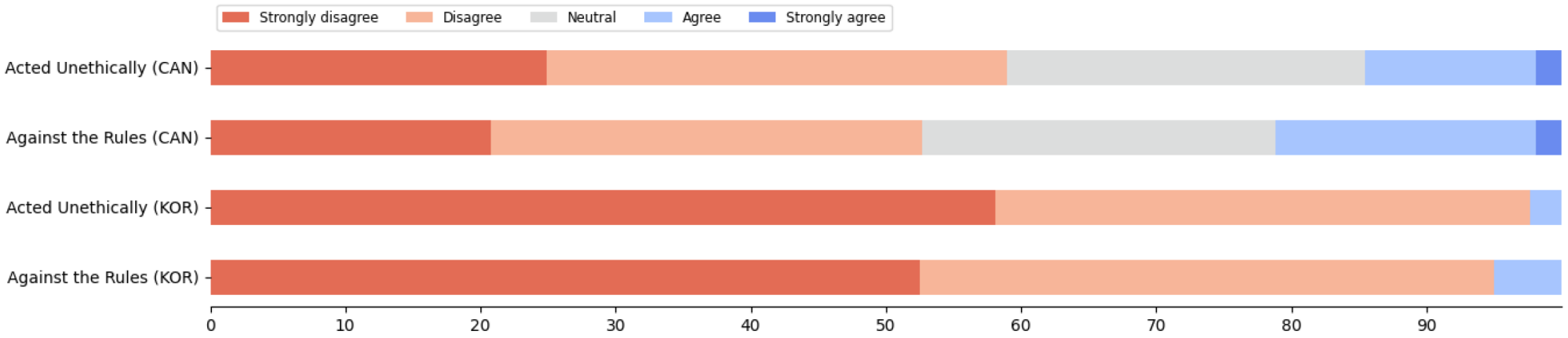

Fig. 11. Responses to Scenario 8. Amount: Low; Understanding: Medium; Stage: Middle.

Table 5. ANOVA Analysis for Scenario Factor Impact on Response to "Acted Unethically" (Top), and "Against the Rules" (Bottom) Prompts

| Factor | SS | MS | Residuals | df | F | p |
|---|---|---|---|---|---|---|
| Amount of generated code | 552.60 | 276.301 | 0.786 | 2 | 351.663 | <**0.0001** |
| Understanding demonstrated | 51.37 | 25.684 | 1.028 | 2 | 24.988 | <**0.0001** |
| Stage of use | 101.41 | 50.704 | 1.004 | 2 | 50.519 | <**0.0001** |
| Amount of generated code | 538.52 | 269.259 | 0.759 | 2 | 354.563 | <**0.0001** |
| Understanding demonstrated | 32.60 | 16.302 | 1.004 | 2 | 16.240 | <**0.0001** |
| Stage of use | 74.47 | 37.235 | 0.984 | 2 | 37.857 | <**0.0001** |

Bold indicates significance after Bonferroni-corrected $\alpha = 0.025$.

eight scenario items. The within-subject factor was Scenario (eight levels), and the between-subject factor was Country (Canada vs. Korea). The ANOVA results can be found in Table 5. All factors are significant ($p < 0.0001$), but the "Amount of generated code in final submission" emerged as the dominant factor influencing student perceptions, with Scenarios 4, 5, and 7 exhibiting the largest mean differences between students in CAN and KOR across both ethicality and rule compliance ratings. This design provided empirical support for analyzing how the degree of AI-generated content influenced cross-cultural ethical evaluations. As shown in Table 2, these three questions were exactly those in the "High" or "Medium" conditions for this factor. This indicates that students at both CAN and KOR were more likely to consider a scenario both unethical and

Table 6. Per-Question Mann–Whitney U Test Results ($U$, $z$, Effect Size, p)

| Scenario and Prompt | $U$ | $z$-score | Effect size | p |
|---|---|---|---|---|
| Scenario 1: Acted unethically | 4,796.5 | 2.41 | 0.1369 | 0.01596 |
| Scenario 1: Against the rules at my institution | 2,997.5 | 4.53 | 0.2573 | **<0.00001** |
| Scenario 2: Acted unethically | 3,244.5 | 4.24 | 0.2408 | **<0.00001** |
| Scenario 2: Against the rules at my institution | 2,679 | 4.56 | 0.259 | **<0.00001** |
| Scenario 3: Acted unethically | 3,908 | 3.98 | 0.226 | **0.00006** |
| Scenario 3: Against the rules at my institution | 2,815.5 | 5.21 | 0.2959 | **<0.00001** |
| Scenario 4: Acted unethically | 2,882.5 | 5.13 | 0.2914 | **<0.00001** |
| Scenario 4: Against the rules at my institution | 2,133.5 | 6.15 | 0.3493 | **<0.00001** |
| Scenario 5: Acted unethically | 2,454 | 6.04 | 0.343 | **<0.00001** |
| Scenario 5: Against the rules at my institution | 1,223.5 | 7.51 | 0.4265 | **<0.00001** |
| Scenario 6: Acted unethically | 3,998.5 | 3.80 | 0.2158 | **0.00014** |
| Scenario 6: Against the rules at my institution | 3,494.5 | 4.34 | 0.2465 | **<0.00001** |
| Scenario 7: Acted unethically | 1,974 | 6.35 | 0.3607 | **<0.00001** |
| Scenario 7: Against the rules at my institution | 1,469.5 | 7.02 | 0.3987 | **<0.00001** |
| Scenario 8: Acted unethically | 2,913.5 | 5.05 | 0.2868 | **<0.00001** |
| Scenario 8: Against the rules at my institution | 2,650.5 | 4.97 | 0.2823 | **<0.00001** |

Bold indicates significance after Bonferroni-corrected $\alpha = 0.0028$.

against institutional rules when it involved the submission of significant amounts of code directly copied from the output of a GenAI tool.

## 5 Discussion

The findings of this study demonstrate significant cross-cultural differences in student perceptions of GenAI use in university coursework between Canadian and Korean students. Consistently across all scenarios, Canadian students were more likely than their Korean counterparts to view Alice's use of GenAI as both unethical and against institutional rules, despite similar institutional policies. This reinforces the argument that cultural frameworks, rather than formal regulations alone, play a critical role in shaping students' ethical judgments regarding emerging technologies in education [13, 67].

Hofstede's cultural dimensions theory provides a useful lens for interpreting these results [16–18]. Canada, characterized by low power distance and high individualism [67], encourages autonomy, personal originality, and strict anti-plagiarism norms [3, 6]. In this context, Canadian students' heightened sensitivity to ethical breaches reflects cultural values that prioritize individual achievement and personal responsibility for academic work. Our findings align with the meta-analysis by Zhao [67], which found that students from individualistic cultures like Canada were more likely to view academic dishonesty, including inappropriate GenAI use, as a serious violation.

In contrast, in Korea, a higher power distance and collectivist orientation [67] frame education around respect for authority, communal knowledge, and adherence to hierarchical structures. More permissive attitudes of Korean students toward AI-assisted work mirror prior findings that collectivist societies often perceive knowledge as a shared resource, and that helping behaviors are seen as morally positive even when viewed as misconduct in Western contexts [33, 61]. Similarly, studies by Park and Kim [42] and Yoon et al. [64] highlight that while Korean students are aware of GenAI's potential, they also navigate its use within a culturally embedded framework that prioritizes conformity and group cohesion.

The stronger correlation observed between perceptions of unethical behavior and perceptions of rule violations ($r = 0.83$, $p < 0.00001$) supported by findings of Cotton et al. [10], who argue that without explicit AI use policies, students tend to project personal ethical standards onto institutional expectations. However, the fact that Canadian students rated rule violations consistently higher than unethical behavior (confirmed by a statistically significant paired $t$-test) suggests a greater internalization of institutional integrity standards, echoing findings by Stone [55] regarding the strong emphasis on individual accountability in North American academic settings.

Our results demonstrate that ethical AI integration in education must be culturally responsive. Bobula [4] and Cotton et al. [10] advocate for clear, structured AI-use guidelines accompanied by training that recognizes both technological and ethical dimensions. However, as Viberg et al. [58] and Tao et al. [56] caution, AI tools themselves often reflect culturally embedded biases, complicating universal ethical frameworks. Our findings suggest that future policy and pedagogical efforts must address these cultural nuances to foster both ethical AI use and academic integrity across diverse educational contexts.

## 6 Conclusion

### 6.1 Study Contributions

This study achieved all three ROs stated in Section 1. First, we successfully examined and compared *student perceptions of the ethicality and compliance with the institutional rules of GenAI use* between Canadian and South Korean university students. The analysis of survey data revealed statistically significant differences across nearly all scenarios, with Canadian students consistently more likely than their South Korean counterparts to judge GenAI-assisted coursework as unethical and against institutional rules, even when the rules were functionally identical.

Second, we analyzed how *the amount and nature of AI-generated content influenced students' ethical judgments across cultural contexts.* ANOVA analysis of the factors contributing to scenarios demonstrated that the amount of AI-generated code included in assignments was the most influential factor shaping students' perceptions. In particular, scenarios involving high or medium levels of copied AI-generated content elicited significantly stronger ethical concerns among all groups of students. The other factors analyzed, including the amount of understanding of code demonstrated, or the stage of development at which the tool was used, were significant, but to a lesser degree.

Third, we interpreted the observed *cross-cultural differences through Hofstede's cultural dimensions framework.* Our findings aligned with established theoretical expectations, namely, Canadian students, representing a low power distance and highly individualistic culture, demonstrated heightened sensitivity to academic integrity breaches. In contrast, South Korean students, shaped by a higher power distance and collectivist orientation, exhibited more permissive attitudes toward AI-assisted learning. This analysis confirms that cultural factors such as individualism, collectivism, and uncertainty avoidance meaningfully mediate students' ethical reasoning regarding GenAI use.

Through these contributions, our study provides better understanding of how cultural context influences ethical perceptions of emerging technologies in higher education and highlights the necessity of culturally responsive AI-use policies.

### 6.2 Implications for Policy and Practice

The findings of this study have several implications for institutional policy development and educational practice. As GenAI tools become increasingly embedded in academic environments, universities must proactively establish governance frameworks that are both culturally responsive and ethically sound. Institutions should consider developing explicit GenAI usage policies that reflect cultural differences in perceptions of collaboration, authority, and academic integrity. For instance, guidelines could clarify distinctions between acceptable assistance and unauthorized content generation, tailored to address varying norms regarding individual achievement and group support. Additionally, culturally sensitive AI literacy programs could be implemented to educate students about ethical AI use, emphasizing not only technical competencies but also critical thinking about academic integrity within different cultural contexts. Finally, regular review and adaptation of academic integrity codes should incorporate specific guidance on the use of GenAI tools, ensuring that policies remain relevant as both technologies and student cultural expectations evolve. Such measures would help institutions support responsible innovation while preserving core educational values across diverse student populations.

### 6.3 Study Limitations

We acknowledge several limitations of our study. First, the study focused on two universities (one in Canada and one in South Korea) which limited the generalizability of the findings to broader national or international student populations. Although the selected institutions provided some cross-cultural comparisons, variations within each country (e.g., between different regions, academic disciplines, or types of institutions) were not captured. Due to the difference in class sizes, the sample sizes of the two universities were also disproportionate. While every effort was made to ensure this was accounted for in the statistical analysis, it should be taken into consideration.

Second, the use of scenario-based surveys, while effective for isolating specific variables, may not fully replicate the complexity of real-world decision-making. Students' actual behavior regarding GenAI use in authentic coursework settings could differ from their self-reported perceptions when responding to hypothetical scenarios.

Additionally, as the study relied on self-reported data, this may be subject to social desirability bias or differing interpretations of survey items across cultural contexts. Triangulation with qualitative interviews or longitudinal studies would strengthen future investigations into the evolving perceptions of GenAI ethics in higher education.

Finally, scenario-level contrasts may partly reflect cross-cultural response-style differences (e.g., neutral-category use), so the Mann–Whitney U test results should not be read as pure location (median) shifts.

### 6.4 Future Research

Future research should build on the findings of this study by expanding the cultural and institutional scope. Comparative studies in a broader range of countries, including those in the Global South and regions with emerging AI adoption, would provide a more comprehensive understanding of how diverse educational and cultural contexts shape perceptions of GenAI ethics. Additionally, longitudinal studies tracking changes in student attitudes over time would help capture how exposure to GenAI tools and evolving institutional policies influence ethical judgments.

Further work could also incorporate mixed-methods approaches, combining scenario-based surveys with qualitative interviews or focus groups, to gain deeper insights into the reasoning processes behind students' ethical evaluations. Investigating the role of academic discipline, level of study, and prior familiarity with AI technologies could offer a more nuanced understanding of subgroup differences within national contexts.

This study lays the groundwork for future replication across different geographic regions, levels of study, and role. In future we hope to replicate this study with a wider variety of institutions from different geographies and cultural contexts, at different levels of academic progress, and also to replicate the study with educators, policy-makers, and academics.

Finally, future studies should critically examine how institutional AI-use policies, when introduced or revised, interact with cultural expectations and individual student values. Such research would support the development of culturally responsive AI governance frameworks that promote both innovation and academic integrity in a rapidly changing educational landscape.

### Data Availability

The data that supports our findings, as well as the full survey instrument, are openly available at https://github.com/BrianHarringtonUTSC/DidAliceDoWrong for any readers who wish to validate or replicate this study.

### Acknowledgments

The authors would like to thank Sohee Kang and Sotirios Damouras of the University of Toronto Scarborough for their help on the statistical analysis of this article.